# Model of Genetic Variation in Human Social Networks

James H. Fowler[1], Christopher T. Dawes[1], Nicholas A. Christakis[2]

[1]*Department of Political Science, University of California, San Diego, CA 92093, USA*

[2]*Department of Health Care Policy, Harvard Medical School, and Department of Sociology, Harvard University, Cambridge, MA 02138, USA*

**Corresponding author: James H. Fowler, <u>jhfowler@ucsd.edu</u>**

**Abstract**

Social networks exhibit strikingly systematic patterns across a wide range of human contexts. While genetic variation accounts for a significant portion of the variation in many complex social behaviors, the heritability of egocentric social network attributes is unknown. Here we show that three of these attributes (in-degree, transitivity, and centrality) are heritable. We then develop a "mirror network" method to test extant network models and show that none account for observed genetic variation in human social networks. We propose an alternative "Attract and Introduce" model with two simple forms of heterogeneity that generates significant heritability as well as other important network features. We show that the model is well suited to real social networks in humans. These results suggest that natural selection may have played a role in the evolution of social networks. They also suggest that modeling intrinsic variation in network attributes may be important for understanding the way genes affect human behaviors and the way these behaviors spread from person to person.

Human social networks are characterized by rich variation at the individual level. Some people have few friends while others have many. Some people are embedded in tightly-knit groups where everyone knows each other, while others belong to many different groups where there is little overlap between friends. To explain this variation, scholars have sought simple models of network formation that generate an empirically realistic distribution of network characteristics as an endogenous outcome of a self-organizing process.

The best-known network formation models start with *identical* individuals that are subjected to social processes that create or exacerbate inequality in a network. For example, in the "scale free" physics model (1) it is the process of growth and, in particular, preferential attachment that drives the "self-organizing" feature of the unequal power-law distribution in the degree. In the economic "connections model" (2-3), individuals who are homogenous *ex ante* endogenously form a star network when actors obtain indirect network benefits and when they are driven by (short-run) economic incentives. And in sociology, actors' preferences for "structural balance" (4) and "homophily" (5) tend to stimulate transitivity in social relationships and the formation of like-minded cliques.

While the *structural processes* in these models generate empirically realistic variation in some network attributes, the effect of *individual characteristics* has been mainly ignored. There have been extensions to the canonical models that do take into account individual heterogeneity (6-12), but these models are usually presented as "robust" versions of the original models, in which the focus still is on the endogenous process (13). In this article, we focus instead on the individual characteristics themselves and explore the possibility that humans are endowed with intrinsic characteristics that affect their network attributes. And our most intrinsic characteristics can be found in our genes.

To test the hypothesis that genes play a role in human social network structures, we use a classic twin study design (14-15). This design measures the heritability of a behavioral trait by comparing trait similarity in (same-sex) monozygotic (MZ) twins who share 100% of their segregating genes to trait similarity in same-sex dizygotic (DZ) twins who share only 50% on average. Under the assumptions of the twin study design, if genetic variation is contributing to variation in the trait, then MZ twins should be significantly more similar than DZ twins. Although some scholars object to the assumptions of the design (see SI), it has been widely used to show that genes play a role in personality (16), intelligence (17-18), and several other behavioral traits (14-15,19-23). In fact, Turkheimer suggests as a "first law of behavior genetics" that all human behavioral traits are heritable (24).

We should therefore not be surprised to learn that individual social network characteristics have a partly genetic basis. However, as we will show, not all network characteristics are significantly heritable, and, more pertinently, specific estimates of heritability can provide a means to test theoretical models of human social networks.

**Results**

The fundamental building blocks of a human social network are *egocentric* properties of each individual in the network: the *degree* (the number of a person's contacts, or *social ties*) and *transitivity* (the likelihood that two of a person's contacts are connected to each other, also called the *clustering coefficient*). A wide variety of social networks can be constructed by altering the distribution of degree and transitivity between individuals (the *nodes* of the network), and these two attributes also have a strong influence on other network characteristics such as *betweenness* *centrality* (the fraction of paths through the network that pass through a given node).  For example, a higher degree is positively correlated with greater centrality.

To measure how much variation in these node-level measures can be attributed to genetic variation, we used an additive genetic model (see SI) to analyze 1,110 twins from a sample of 90,115 adolescents in 142 separate school friendship networks in the National Longitudinal Study of Adolescent Health (the *Add Health* study -- see SI for description).  The results show that genetic factors account for 46% (95% C.I. 23%, 69%) of the variation in in-degree (how many times a person is named as a friend), but heritability of out-degree (how many friends a person names) is not significant (22%, C.I. 0%, 47%).  In addition, node transitivity is significantly heritable, with 47% (C.I. 13%, 65%) of the variation explained by differences in genes.  We also find that genetic variation contributes to variation in other network characteristics; for example, betweenness centrality is significantly heritable (29%, C.I. 5%, 39%).

These results allow us to reject the hypothesis that genes have no effect on human social networks.  However, they also focus our attention on what kinds of attributes are heritable.  For example, it is striking that in-degree is significantly heritable while out-degree appears not to be.  There are many potentially interesting causal pathways from genes to human network structure that merit exploration.  For example, it was recently shown that the -G1438A polymorphism within the promoter region of the 5-HT2A serotonin receptor gene is associated with variation in popularity (25).  However, here we focus on the important *implications* of such variation -- whatever its specific genetic determinant -- for models of human social networks.

Network models that do not include intrinsic node characteristics cannot generate heritability in network attributes.  The reason is that nodes without their own individual properties can be interchanged without affecting the structure of the network (6). Likewise, genes give people their individuality; without genetic variation, human characteristics cannot, by definition, be heritable. Thus, to generate heritability in a model of human social networks, nodes must be endowed with characteristics that actually exhibit variation, and these characteristics must be associated with node network measures.

We surveyed the existing literature for network models that incorporate intrinsic node characteristics. "Hidden variables" models incorporate variation in an attribute regulating the formation of social ties (6). For example, a "fitness" parameter has been used to explore the conditions under which a late entrant might dominate networks constructed via preferential attachment (7,26). This "fitness" model was in fact proposed in order to take into account the fact that some nodes are intrinsically more attractive. In an alternative model (8), nodes are placed in "social space" (9) or a "latent space" (10) where greater social distance reduces the likelihood of a social tie. The social space model (8) in particular generates three outcomes that are characteristic of human social networks (as distinct from technological or biological networks): high transitivity, positive degree-degree correlation (popular people have popular friends), and community structure within the network (11). Finally, exponential random graph models (ERGMs) are statistical network characterizations that can incorporate node heterogeneity to explain degree heterogeneity and population-level average transitivity (12).

We developed a "mirror network" method to test whether the "fitness," (7) "social space," (8) "ERGM," (12) or regular Erdos-Renyi "random" network (27) models generate heritability in degree or transitivity (see SI). In this method, we create one set of nodes with intrinsic characteristics drawn from a probability distribution as defined by each of the models. We then follow the procedures outlined in the network model being tested for connecting nodes. Once that is complete, we create a second set of nodes with intrinsic characteristics drawn from the same probability distribution as before. We randomly choose one node from the first set and copy its characteristics to one randomly chosen node in the second set to create a pair of "twins." We then follow the procedures outlined in the network model being tested for connecting the second set of nodes to create a "mirror network." This is like creating $N$ identical twin pairs and putting one twin from each pair in two separate environments. The initial randomization ensures that twins have uncorrelated environments prior to the onset of edge formation. Therefore, any resulting correlation in a twin pair's network measures (or the network measures of their friends) is an outcome of the edge-generation process, not the other way around.

Once the two networks have been independently constructed, we record relevant network measures for each of the two twins (in-degree, out-degree, transitivity, and betweenness centrality). We then repeat this procedure 10,000 times. The Pearson correlation between the twins gives an estimate of the proportion of the variation of the network measure that is explained by intrinsic node characteristics, analogous to the phenotypic variance explained by genes in models of identical twins reared apart (16). The "mirror network" method rejects all extant models of social network formation because they do not generate heritability that falls within the confidence intervals of the empirical estimates (Fig.1). The ERGM comes closest, generating realistic heritability in in-degree and betweenness centrality, but it does not generate realistic heritability in transitivity.

We therefore developed an alternative "Attract and Introduce" model (see SI) built on two assumptions. First, some individuals are inherently more attractive than others, whether physically or otherwise, so they receive more friendship nominations. Second, some individuals are inherently more inclined to introduce new friends to existing friends (and hence these individuals will indirectly enhance their node transitivity).  In the "Attract and Introduce" model, individuals are chosen randomly to form ties and introduce their friends until a fixed number of ties for the whole network is reached (the alternative models either follow the same rule or they establish probabilities of tie formation that yield a fixed number of ties in expectation for a given network size).  The model has just two parameters, one controlling the distribution of $p_{attract}$ that is the probability of being named as a friend, and one controlling the distribution of $p_{introduce}$ that is the probability of introducing one's friends to each other.

The "Attract and Introduce" model generates heritability for in-degree, transitivity, and betweenness centrality that falls within the range of heritability observed in the real data (Fig.1). The model also yields other important characteristics of human networks.  Fig. 2 shows that the tail of the degree distribution falls between the straight line of a power-law distribution (as generated by the fitness model) and the fast cutoff of an exponential distribution (as generated by the social space and Erdos-Renyi models) (Fig. 2).  The "Attract and Introduce" model also generates positive degree-degree correlation ($\rho = 0.18$), high transitivity ($c=0.18$), a relationship between node degree and transitivity that closely follows the observed relationship in the Add Health data (Fig.3), and realistic community structure with significant modularity (Fig.4) (24). Finally, our model also generates motif structures that have a higher likelihood than all the proposed alternatives (Fig.5).  These motif structures are patterns of ties in sets of 3 nodes or sets of 4 nodes, and their frequency creates a network "fingerprint" that can be used to identify a unique set of observed or simulated networks (see SI).

**Discussion**

To date, there has been relatively little attention to the role of individual heterogeneity in the formation of social networks. The evidence we present here suggests that egocentric properties are significantly heritable in human social networks. It is therefore important to make individual characteristics just as focal in the modeling of social networks as structural processes are. Although it may not be surprising that genetic variation influences network formation, the effects are large enough that it is hard to argue that they can be ignored. Our "Attract and Introduce" model accounts for the role genes play not only in direct relationships (in-degree) but also in indirect relationships (transitivity and centrality), and as a consequence it is able to generate realistic large-scale community structure. We hope that this approach will generate broad interest in modeling individual heterogeneity and in using methods like the mirror network technique to test future models of network formation.

In the "Attract and Introduce" model, genes shape networks, but it is also may be the case that networks shape genes. Scholars studying the evolution of cooperation in humans have recently turned their attention to the structure of social networks underlying human interactions (28). For example, in a fixed social network, cooperation can evolve as a consequence of "social viscosity" even in the absence of reputation effects or strategic complexity (29-30). Different network structures can speed or slow selection and in some cases they completely determine the outcome of a frequency-dependent selection process (31). Moreover, adaptive selection of network ties by individuals on evolving graphs can also influence the evolution of behavioral types (32-34). This research provides several theoretical examples of how natural selection can yield stable variation in local network structures. Future work should explore whether social networks may also result from (or contribute to) other sources of genetic variation in humans and other species, such as life history tradeoffs (35), balance between mutation and selection (36), sexually antagonistic selection (37), or the search for desirable partners also sought by others.

There may be many reasons for genetic variation in the ability to attract or the desire to introduce friends.  More friends may mean greater social support in some settings or greater conflict in others.  Having denser social connections may improve group solidarity, but it might also insulate a group from beneficial influence or information from individuals outside the group.  And while it is possible that variation in individual social network attributes is incidental to natural selection processes operating on other traits, it is remarkable that network traits have significant heritability.  Another area of future research should be the identification of mediating mechanisms like personality traits and the specific genes that may be involved.

Finally, social networks may serve the adaptive (or maladaptive) function of being a vehicle for the transmission of emotional states, material resources, or information (e.g., about resource or partner availability) between individuals (38).  Some traits that appear to spread in social networks also appear to be heritable (such as obesity (20,39), smoking behavior (40,41), happiness (42,43), and even political behavior (44-47)), suggesting that a full understanding of these traits may require a better understanding of the genetic basis of social network topology.  To that end, we urge network theorists, behavior geneticists, evolutionary biologists, and social scientists to unify their theories regarding the structure and function of social networks, and their genetic antecedents.

**Materials and Methods**

For the "Attract and Introduce" model we assume there are $N$ nodes and $E$ edges. Each node is permanently endowed with two characteristics, $p_{\text{attract}}^{j}$ and $p_{\text{introduce}}^{i}$, with values randomly drawn from fixed distributions. The distribution of $p_{\text{attract}}^{j}$ is based on a single parameter $\alpha \in [0,1]$ such that $\Pr\left(p_{\text{attract}}^{j} \sim \text{Uniform}[0,1]\right) = \alpha$ and $\Pr\left(p_{\text{attract}}^{j} = 0\right) = 1 - \alpha$. The distribution of $p_{\text{introduce}}^{i}$ is based on a single parameter $\beta \in [0,1]$ such that $\Pr\left(p_{\text{introduce}}^{j} = 1\right) = \beta$ and $\Pr\left(p_{\text{introduce}}^{j} = 0\right) = 1 - \beta$.

At each time period, nodes $i$ and $j$ are randomly chosen from the population, and with probability $p_{\text{attract}}^{j}$ a social tie from $i$ to $j$ forms. If this occurs, then with probability $p_{\text{introduce}}^{i}$, $i$ chooses to introduce $j$ to all of his "friends" (the other nodes to which $i$ already sends a tie). If $i$ does introduce, then each friend sends a tie to $j$ with probability $p_{\text{attract}}^{j}$ and $j$ sends a tie to each friend with probability $p_{\text{attract}}^{k}$ that corresponds to each $k$th friend. This process is repeated until at least $E$ ties are generated. In the Supporting Information we show code used to generate this model and test it using the "mirror network" method to assess heritability.

To establish the best fitting distributions for $p_{\text{attract}}^{j}$ and $p_{\text{introduce}}^{i}$, we optimized one parameter for each to generate the empirically-observed average transitivity and heritability of in-degree and node transitivity in a network where $N$=750 and $E$=3150 (reflecting the typical school network in the Add Health data). The results of this optimization yielded $\alpha \approx 0.9$ and $\beta \approx 0.3$. The distribution of attractiveness that fit the data suggests about one-tenth of the individuals have very low attractiveness while the remaining nine-tenths are approximately evenly distributed between low, medium, and high attractiveness. The probability a person will have the desire to introduce is about three-tenths.

**Acknowledgements:**  *Research supported by National Institute on Aging grant P-01 AG-031093 and National Science Foundation grant SES-0719404.*

## References

1. Barabasi, A.-L. and R. Albert (1999), "Emergence of scaling in random networks", Science, 286, pp. 509-512.

2. Jackson, M.O. and A. Wolinsky (1996), "A strategic model of economic and social networks", Journal of Economic Theory, 71, pp. 44-74.

3. Bala, V. and S. Goyal (2000), "A non-cooperative model of network formation", Econometrica, 68(5), 1181-1231.

4. Cartwright, D. and F. Harary (1956), "Structural balance: a generalization of Heider's theory", Psychological Review, 63, pp. 277-292.

5. McPherson, M., L. Smith-Lovin, and J.M. Cook (2001), "Birds of a feather: Homophily in social networks", Annual Review of Sociology, 27, pp. 415-444.

6. Park J.Y., Barabási, A.L. Distribution of Node Characteristics in Complex Networks. *PNAS* 104: 17916-17920 (2007)

7. Bianconi, G., Barabási, A.L. Competition and multiscaling in evolving networks. *Europhys. Lett.* 54: 436 (2001).

8. Boguñá, M., Pastor-Satorras, R., Díaz-Guilera, A., Arenas, A. Models of social networks based on social distance attachment. *Phys Rev E* 70: 056122 (2004).

9. Watts, D.J., Dodds, P.S., Newman, M.E.J. Identity and Search in Social Networks. *Science* 296: 1302 (2002).

10. Hoff, P.D., Raftery, A.E., Handcock, M.S. Latent Space Approaches to Social Network Analysis. *Journal of the American Statistical Association* 97:1090-1099 (2002)

11. Girvan, M., Newman, M.E.J. Community structure in social and biological networks. *PNAS* 99: 7821 (2002).

12. Snijders, T.A.B., Pattison, P.E., Robins, G.L., Handcock, M.S. New specifications for exponential random graph models. *Sociological Methodology* 36: 99-153 (2006)

13. Boccaletti, S., Latora, V., Moreno, Y., Chavez, M., Hwang, D.U. Complex networks: Structure and dynamics. *Physics Reports* 424: 175 (2006).

14. Evans, D.M., Gillespie, N.A., Martin, N.G. Biometrical Genetics. *Biological Psychology* 61, 33 (2002).

15. Neale, M.C., Cardon, L.R. *Methodology for Genetic Studies of Twins and Families. Dordrecht*, The Netherlands: Kluwer (1992).

16. Bouchard, T.J., Lykken, D.T., McGue, M., Segal, N.L., Tellegen, A. Sources of human psychological differences: the Minnesota Study of Twins Reared Apart. *Science* 250: 223 (1990).

17. Plomin, R. Genetics and general cognitive ability. *Nature* 402: C25-C29 (1999).

18. Devlin, B., Daniels, M., Roeder, K. The heritability of IQ. *Nature* 388: 468-471 (1997)

19. Defries, J.C., Fulker, D.W., Labuda, M.C. Evidence For A Genetic Etiology In Reading-Disability Of Twins. *Nature* 329: 537-539 (1987)

20. Herbert A., Gerry, N.P., McQueen, M.B., et al. A common genetic variant is associated with adult and childhood obesity. *Science* 312: 279-283 (2006)

21. Fox, P.W., Hershberger, S.L., Bouchard, T.J. Genetic and environmental contributions to the acquisition of a motor skill. *Nature* 384: 356-358 (1996)

22. Fowler, J.H., Baker, L.A., Dawes, C.T. Genetic Variation in Political Participation. *American Political Science Review* 102: 233-248 (2008)

23. Settle, J.E. Dawes, C.T., Fowler, J.H. The Heritability of Partisan Attachment. *Political Research Quarterly*, in press (2009)

24. Turkheimer, E. Three Laws of Behavior Genetics and What They Mean. *Current Directions in Psychological Science* 9: 160-164 (2000)

25. Burt, S.A. Genes and popularity: Evidence of an evocative gene-environment correlation. *Psychological Science* 19: 112-113 (2008)

26. Kong, J.S., N. Sarshar, and V.P. Roychowdhury (2008), "Experience versus talent shapes the structure of the Web", *PNAS*, 105(37), 13724-13729.

27. Erdős, P.; Rényi, A. On Random Graphs. I. *Publicationes Mathematicae* 6: 290-297 (1959)

28. Szabo G, Fath G. Evolutionary games on graphs. *Physics Reports* 446: 97-216 (2007)

29. Ohtsuki H, Hauert C, Lieberman E, et al. A simple rule for the evolution of cooperation on graphs and social networks. *Nature* 441, 502-505 (2006)

30. Nowak, M.A., Five rules for the evolution of cooperation. *Science* 314: 1560-1563 (2006)

31. Lieberman E, Hauert C, Nowak MA. Evolutionary dynamics on graphs. *Nature* 433, 312-316 (2005)

32. Pacheco JM, Traulsen A, Nowak MA. Coevolution of strategy and structure in complex networks with dynamical linking. *Physical Review Letters* 97: 258103 (2006)

33. Santos FC, Pacheco JM, Lenaerts T. Cooperation prevails when individuals adjust their social ties. *PLOS Computational Biology* 2: 1284-1291

34. Skyrms, B., Pemantle, R. A dynamic model of social network formation. *PNAS* 97:9340-9346 (2000).

35. Wolf, M., van Doorn, G.S., Leimar, O., et al. Life-history trade-offs favour the evolution of animal personalities. *Nature* 447: 581-584 (2007)

36. Barton, N. H. & Keightley, P. D. Understanding quantitative genetic variation. *Nature Rev. Genet.* 3, 11–21 (2002)

37. Foerster, K., Coulson, T., Sheldon, B.C., Pemberton J.M., Clutton-Brock, T.H., Kruuk, L.E.B. Sexually antagonistic genetic variation for fitness in red deer. *Nature* 447: 1107-1110

38. Gervais, M., Wilson, D.S. The evolution and functions of laughter and humor: a synthetic approach. *Quarterly Review of Biology* 80: 395-430 (2005)

39. Christakis, N.A., Fowler, J.H. The Spread of Obesity in a Large Social Network Over 32 Years. *New England Journal of Medicine* 357: 370-379 (2007)

40. Carmelli, D., Swan, G.E., Robinette, D. et al. Genetic Influence On Smoking - A Study Of Male Twins. *New England Journal of Medicine* 327: 829-833 (1992)

41. Christakis, N.A., Fowler, J.H. The Collective Dynamics of Smoking in a Large Social Network. *New England Journal of Medicine* 358: 2249-58 (2008)

42. Lykken, D., Tellegen, A. Happiness is a Stochastic Phenomenon. *Psychological Science* 7: 186-189 (1996)

43. Fowler, J.H., Christakis, N.A. The Dynamic Spread of Happiness in a Large Social Network. *Bristish Medical Journal* 337: TBA (2008).

44. Fowler, J.H., Schreiber, D. Biology, Politics, and the Emerging Science of Human Nature. *Science* 322: 912-914 (2008)

45. Nickerson, D.W. Is Voting Contagious? Evidence from Two Field Experiments, *American Political Science Review* 102: 49-57 (2008)

46. Fowler, J.H., Baker, L.A., Dawes, C.T. Genetic Variation in Political Participation. *American Political Science Review* 102: 233-248 (2008)

47. Fowler, J.H., Dawes, C.T. Two Genes Predict Voter Turnout. *Journal of Politics* 70: 579-594 (2008)

48. Middendorf, M., Ziv, E., Wiggins, C.H. Inferring network mechanisms. *PNAS* 102: 3192-3197 (2005).

49. Milo, R., Shen-Orr, S., Itzkovitz, S., Kashtan, N., Chklovskii, D., Alon, U. Network Motifs: Simple Building Blocks of Complex Networks. *Science* 298: 824-827 (2002)

50. Holland, P., Leinhardt, S. in *Sociological Methodology*, D. Heise, Ed. (Jossey-Bass, San Francisco, 1975), pp. 1-45.

51. Wasserman, S., Faust, K. *Social Network Analysis.* Cambridge Univ. Press, New York, (1994)

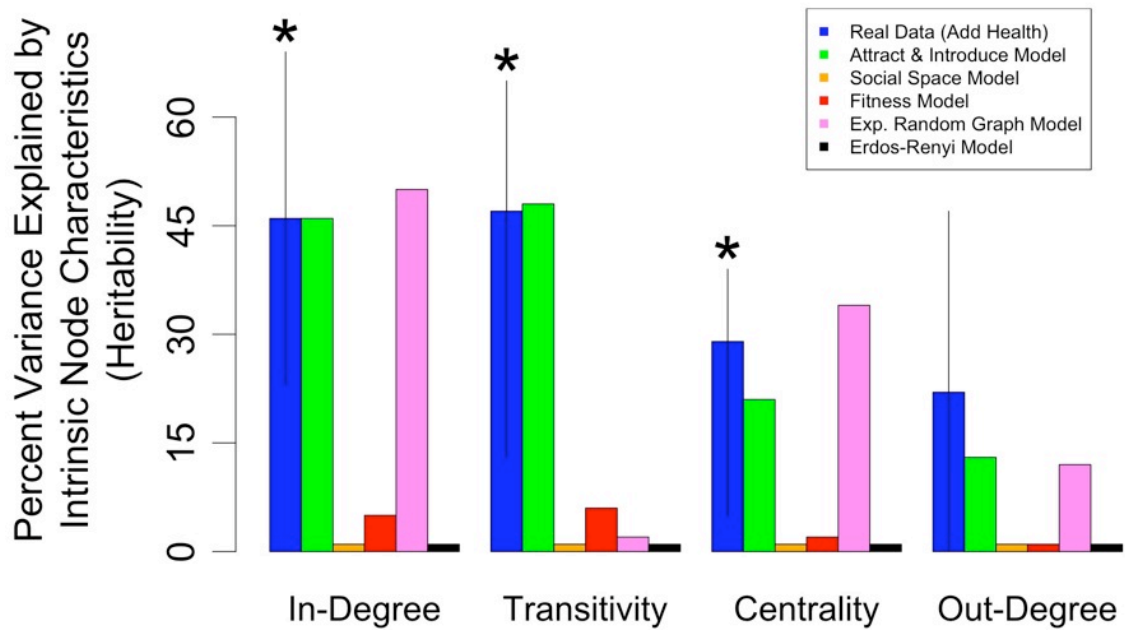

**Fig. 1.** Heritability of network characteristics in a real social network (Add Health) and simulated networks based on five models, the "Attract and Introduce" model; a "social space" model (8); a "fitness" model (7); an "ERGM", or exponential random graph model (12); and an Erdos-Renyi random network (27). We used additive genetic models of monozygotic and dizygotic twins in Add Health to measure the heritability of network characteristics in real human social networks (see SI). The blue bars show that genetic variation accounts for significant variation in in-degree, transitivity, and betweenness centrality (vertical lines indicate 95% confidence intervals, asterisks indicate which confidence intervals exclude 0). These results suggest intrinsic characteristics have an important impact on the fundamental building blocks of real human social networks. Heritability of out-degree is not significant. To see which network models are capable of generating heritability consistent with the empirical observations, we simulated 1000 pairs of networks and used the "mirror network" method for each proposed model to measure how much variance in network measures can be explained by intrinsic node characteristics. Compared to heritability estimates from the real social network data, all proposed models are rejected because they fall outside the confidence intervals except ERGM for in-degree and "Attract and Introduce" for all three significantly heritable network properties (in-degree, transitivity, and centrality).

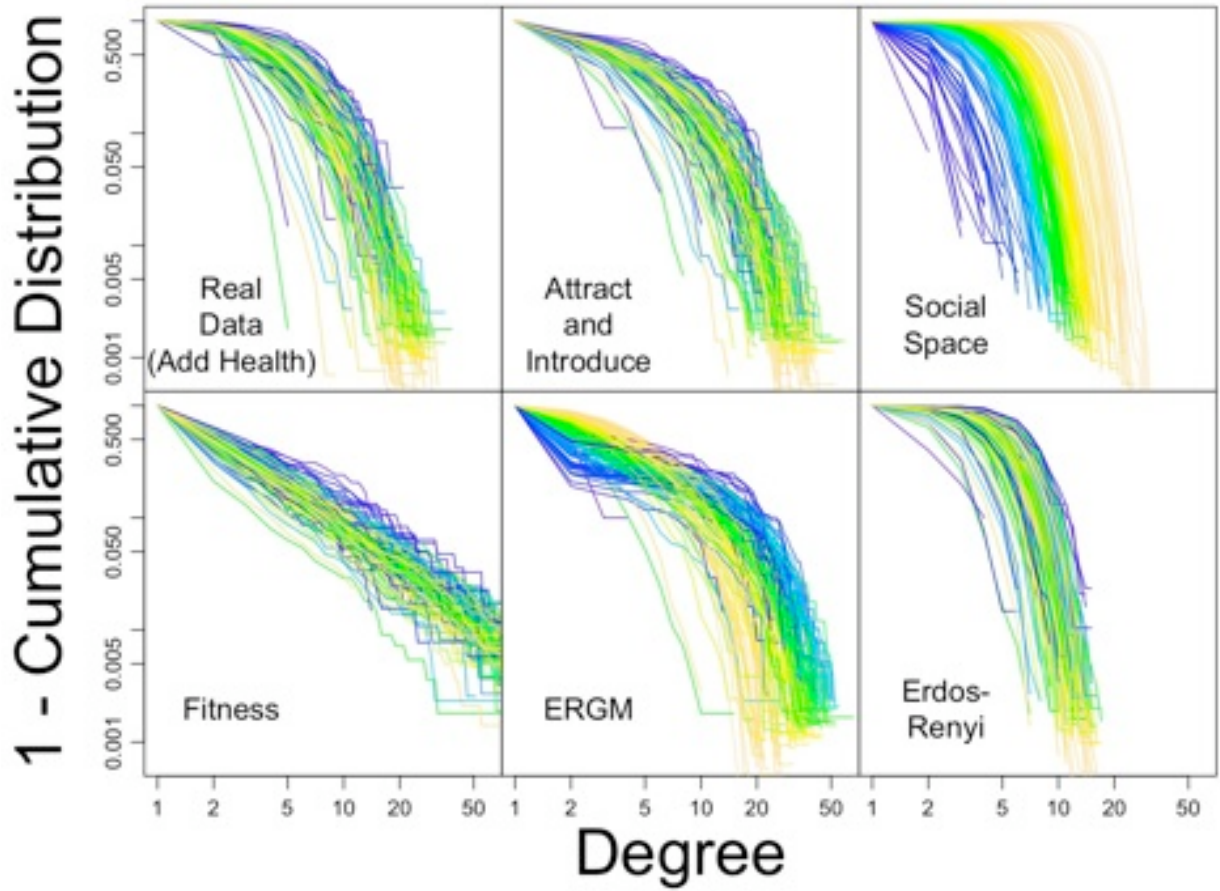

**Fig. 2.** Comparison of degree distributions in real social networks (within each of the 146 schools in the Add Health sample) and like-sized simulated networks based on five models, the "Attract and Introduce" model; a "social space" model (8); a "fitness" model (7); an "ERGM", or exponential random graph model (12); and an Erdos-Renyi random network (27). In the upper left panel, each line indicates the in-degree distribution for each school in Add Health. In the remaining panels, each line indicates the degree distribution in one simulation that assumes the same number of nodes and edges as each of the 146 schools. The color of each line indicates the size of the network (number of nodes) with yellow shades for small, green for medium, and blue for large networks (total range 9 to 2724, mean 752). The fitness model generates a power-law tail in the degree distribution that is overdispersed (more nodes with higher degree, fewer nodes with lower degree) compared to the real data, while the Erdos-Renyi and social space models generate an exponential cutoff and a degree distribution that is underdispersed. Both ERGM and the "Attract and Introduce" model are slightly overdispersed, but ERGM in particular shows greater dispersion for larger networks (blue lines are lower for low degree and higher for high degree) in a pattern that does not exist in the real data (the social space model also exhibits an unrealistically strong relationship between network size and dispersion). The "Attract and Introduce" model produces variation in degree distributions across networks that is similar to Add Health.

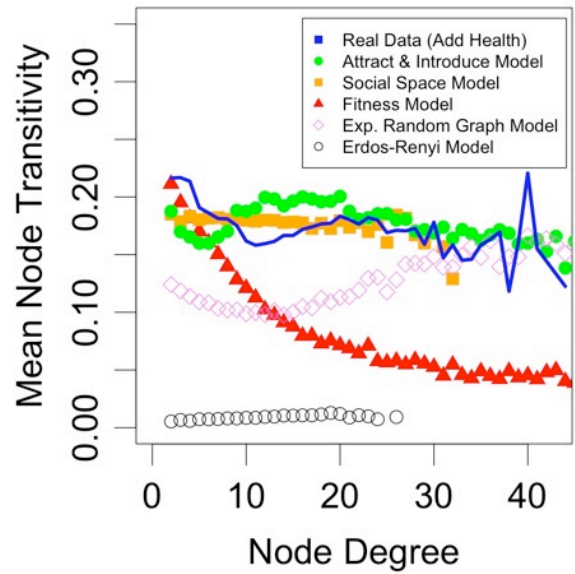

**Fig. 3.** Comparison of node degree and mean node transitivity in a real social network (Add Health) and simulated networks based on five models, the "Attract and Introduce" model; a "social space" model (8); a "fitness" model (7); an "ERGM", or exponential random graph model (12); and an Erdos-Renyi random network (27). We used the number of nodes and edges for each observed network in Add Health to generate 1000 simulated networks for each proposed model and then calculated the mean node transitivity for all nodes of a given degree. The "Attract and Introduce" model deviates least from the observed data.

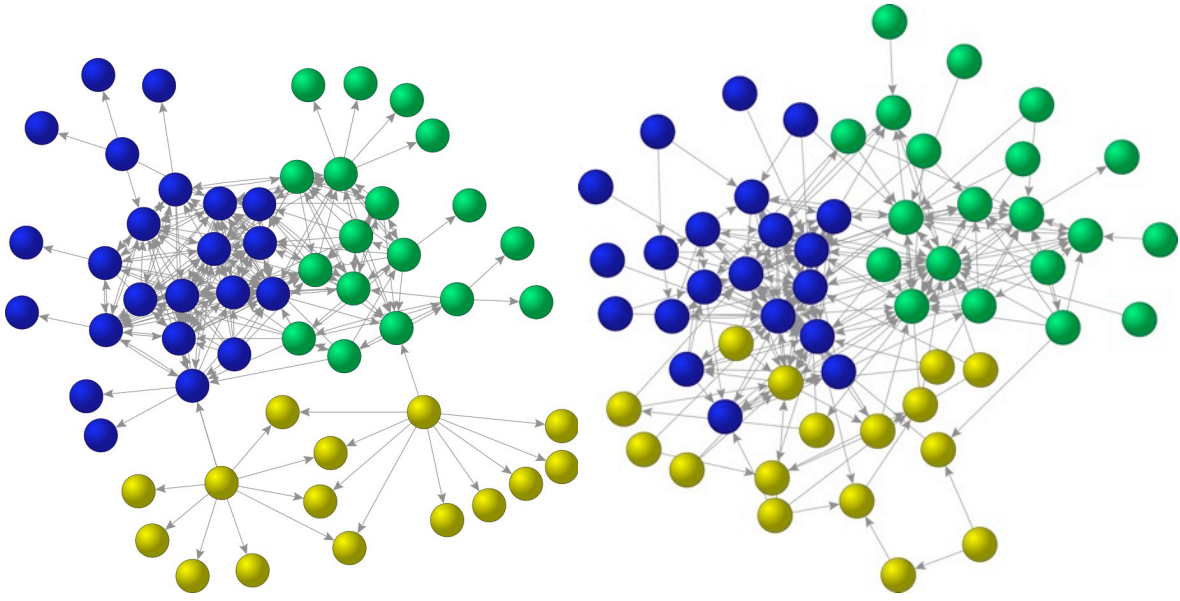

**Fig. 4. (Left)** School 115 (57 nodes and 252 ties) from the Add Health data. **(Right)** Simulated network using the "Attract and Introduce" model and the same number of nodes and ties. These networks show significant modularity with well-defined communities that have many connections within their group and few connections to other groups. Colors indicate communities that maximize modularity (11) (modularity = 0.35 in real network; modularity = 0.34 in simulated network).

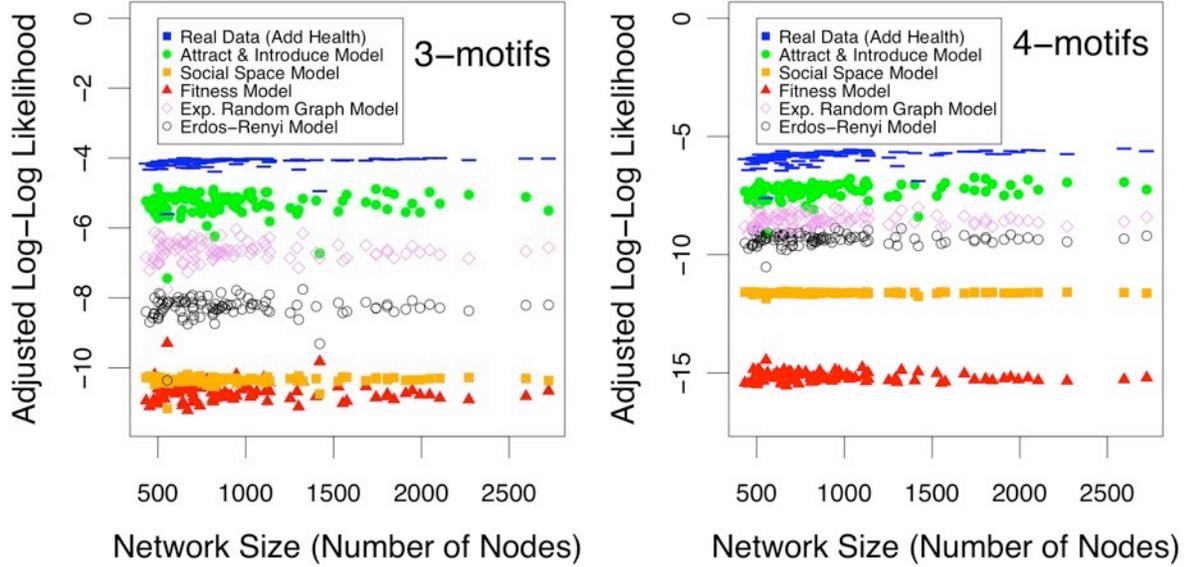

**Fig. 5.** The "Attract and Introduce" model best fits the motif structure of observed networks. **(Left)** We simulate 100 networks from each proposed model using the empirical distribution of nodes and edges in each Add Health network. We then count the total number of 3-motifs (isomorphic combinations of ties that connect 3 nodes) in each network and divide by the total number of motifs in that network to generate the empirical probability that three nodes form any given motif (the 3-motif "fingerprint" of the network) (48-51). For each motif, we fit one dimension of a multivariate beta density across all simulated networks to characterize the empirical probability of observing a given motif for a given model. We then use this estimated density to calculate the likelihood that the observed network could have been generated from the proposed model. For ease of exposition, we show adjusted likelihoods $-\log(c - LL)$, where $c$ is a constant across all networks and models and $LL$ is the log likelihood of generating an observed network. Each point in the figure represents the adjusted likelihood that a proposed model generates an observed Add Health network of a given size. Among all proposed models, "Attract and Introduce" (green circles) is the most likely to generate the 3-motif fingerprint in all of the Add Health networks, as shown. **(Right)** We repeat the fingerprint procedure for 4-motifs (isomorphic combinations of ties that connect 4 nodes). "Attract and Introduce" is also the most likely to generate the 4-motif fingerprint in 98 of 100 Add Health networks. Exponential random graph models (violet diamonds) are most likely to generate the 4-motif fingerprint for 2 of 100 Add Health networks. For comparison, we also show adjusted likelihoods generated using a multivariate beta density fit to the actual data (blue dashes). In the Supporting Information we show that simulated networks are classified correctly in 10,000 out of 10,000 tests using the fingerprint procedure.

**Supporting Information for "A Model of Genetic Variation in Human Social Networks"**

James H. Fowler, Christopher T. Dawes, Nicholas A. Christakis

**The Add Health Data**

The National Longitudinal Study of Adolescent Health[1] (*Add Health*) is a study that, among other topics, explores the causes of health-related behavior of adolescents in grades 7 through 12 and their outcomes in young adulthood. Three waves of the Add Health study have been completed: Wave I was conducted in 1994-1995, Wave II in 1996, and Wave III in 2001-2002.

In Wave I of the Add Health study, researchers collected an "in-school" sample of 90,118 adolescents chosen from a nationally-representative sample of 142 schools. These students filled out questionnaires about their friends, choosing up to 5 male and 5 female friends who were later identified from school-wide rosters to generate information about each school's complete social network. We use these nominations to measure the *in-degree* (the number of times an individual is named as a friend by other individuals) and *out-degree* (the number of individuals each person names as a friend) of each subject. The in-degree is virtually unrestricted (the theoretical maximum is $N-1$, the total number of other people in the network) but the out-degree is restricted to a maximum of 10 due to the name generator used by Add Health. Fortunately, most subjects (90.0%) named fewer than the maximum, and there is substantial variation in the total number of friends named by each person (mean = 3.8, standard deviation = 3.7).

We also measure *transitivity* as the empirical probability that any one of an individual's friends names or is named by a friend by any of that individual's other friends. This is just the total number of triangles of ties divided by the total possible number of triangles for each individual.

Finally, we measure *betweenness centrality* (Freeman 1977) which identifies the extent to which an individual in the network is critical for passing support or information from one individual to another. If we let $\sigma_{ik}$ represent the number of shortest paths from subject $i$ to subject $k$, and $\sigma_{ijk}$ represent the number of shortest paths from subject $i$ to subject $k$ that pass through subject $j$, then the betweenness measure $x$ for subject $j$ is $x_j = \sum_{i \neq j \neq k} \frac{\sigma_{ijk}}{\sigma_{ik}}$.

Note that for the purpose of measuring transitivity and betweenness centrality, we assume all directed ties are undirected, so that a tie in either direction becomes a mutual tie. For example, we consider the case where A names B, B names C, and C names A to be transitive. Likewise, if A names B, A names C, and B names C, we consider the relationships to be transitive for all three individuals.

The Add Health team created a genetically informative sample of sibling pairs, including all adolescents that were identified as twin pairs, half-siblings, or unrelated siblings raised together. Twins and half biological siblings were sampled with certainty. The Wave I sibling-pairs sample has been found to be similar in demographic composition to the full Add Health sample.[2]

**The Twin Study Method**

In order to estimate the heritability of egocentric social network attributes, we study the patterns of same-sex (identical) monozygotic (MZ) twins who were conceived from a single fertilized egg and same-sex (non-identical) dizygotic (DZ) twins who were conceived from two separate eggs. MZ twins share 100% of their segregating genes, while DZ twins share only 50% on average. Thus, if network attributes are heritable, MZ twins should exhibit more similarity than DZ twins. Moreover, if it is assumed that MZ twins and DZ twins share comparable environments, then we can use these concordances to estimate explicitly the proportion of the overall variance attributed to genetic, shared environmental, and unshared environmental factors. Very few differences have been found between twins and non-twins, therefore we expect the results for twins to be generalizable to a non-twin population.[3]

Some scholars have objected to the assumption that MZ and DZ environments are comparable, arguing that the identical nature of MZ twins cause them to be more strongly affiliated and more influenced by one another than their non-identical DZ counterparts. If so, then greater concordance in MZ twins might merely reflect the fact that their shared environments cause them to become more similar than DZ twins. However, studies of twins raised together have been validated by studies of twins reared apart,[4] suggesting that the shared environment does not exert enhanced influence on MZ twins. More recently, Visscher et al. utilize the small variance in percentage of shared genes among DZ twins to estimate heritability without using any MZ twins, and they are able to replicate findings from studies of MZ and DZ twins reared together.[5] Moreover, personality and cognitive differences between MZ and DZ twins persist even among twins whose zygosity has been miscategorized by their parents,[6] indicating that being mistakenly treated as an identical twin by one's parents is not sufficient to generate the difference in concordance. And, although MZ twins are sometimes in more frequent contact with each other than DZ twins, it appears that twin similarity (e.g., in attitudes and personality) may cause greater contact rather than vice versa.[7] Finally, contrary to the expectation that the influence of the unshared environment would tend to decrease concordance over time, once twins reach adulthood, MZ twins living apart tend to become more similar with age.[6]

The Add Health data has been used in a wide variety of twin studies.[8] As a result, there have been several analyses of the comparable environments assumption for MZ and DZ twins. One of these studies reported that the environments were not comparable,[9] but other scholars have pointed to serious deficiencies in this work that negated its conclusions.[10] For example, Horwitz, et al.[9] showed that including observed social variables in a twin model causes the *p*-value on the genetic component for males trying alcohol to change from being just below 0.05 to just above it. Freese and Powell[10] note that this is unsurprising since adding variables to a regression can have a substantial effect on efficiency. Even worse, they point out that Horwitz, et al.[9] do not acknowledge that their own fit statistics indicate the models with and without social variables are statistically indistinguishable, suggesting that the model with additional variables should be rejected.

The twin study design has been used frequently to identify the relative degree to which genetic and environmental factors influence an observed outcome.[11-12] The basic twin model assumes that the variance in observed behavior can be partitioned into additive genetic factors (A), and environmental factors which are shared or common to co-twins (C), and unshared environmental (E). This is the so-called ACE model. The role of genotype and environment are not measured directly but their influence is inferred through their effects on the covariances between twin siblings.[12] No observed covariates are needed in the model because the degree to which they contribute to variance is a part of one of three variance components (A, C, and E). More formally, these components are derived from known relationships between three observed statistics[11]:

$$\sigma_P^2 = \sigma_A^2 + \sigma_C^2 + \sigma_E^2$$

$$COV_{MZ} = \sigma_A^2 + \sigma_C^2$$

$$COV_{DZ} = 0.5\sigma_A^2 + \sigma_C^2$$

In this equation, $\sigma_P^2$ is the observed phenotypic variance (the same for monozygotic and dizygotic twins), $COV_{MZ}$ and $COV_{DZ}$ are the observed covariances between monozygotic twins who share all their genes and dizygotic twins who share only half on average, and $\sigma_A^2$, $\sigma_C^2$, and $\sigma_E^2$ are the variance components for genes, common environment, and unshared environment, respectively. This is a system of three equations and three unknowns so it is identified:

$$\begin{pmatrix} \sigma_A^2 \\ \sigma_C^2 \\ \sigma_E^2 \end{pmatrix} = \begin{pmatrix} 1 & 1 & 0 \\ 0.5 & 1 & 0 \\ 1 & 1 & 1 \end{pmatrix}^{-1} \begin{pmatrix} COV_{MZ} \\ COV_{DZ} \\ \sigma_P^2 \end{pmatrix}$$

Heritability, or the proportion of the variance explained by genetic factors, can be estimated as $\sigma_A^2/(\sigma_A^2 + \sigma_C^2 + \sigma_E^2)$. We use the software package MX to estimate this structural equations model.[13] Table S1 shows the complete results for each of the social network measures of interest.

### Table S1: ACE Model Estimates from the Add Health Data

| Model | *Proportion of Variance Explained by* | | | |
|---|---|---|---|---|
| | Genetic Factors | Common Environment | Unshared Environment | Model Fit (-2LL) |
| *In-Degree* | 0.46 (0.23, 0.69) | 0.21 (0.00, 0.40) | 0.34 (0.28, 0.40) | 2386.11 |
| *Transitivity* | 0.47 (0.13, 0.65) | 0.09 (0.00, 0.36) | 0.44 (0.35, 0.56) | 2033.91 |
| *Betweenness Centrality* | 0.29 (0.05, 0.39) | 0.00 (0.00, 0.19) | 0.71 (0.61, 0.81) | 2489.30 |
| *Out-degree* | 0.22 (0.00, 0.47) | 0.16 (0.00, 0.40) | 0.63 (0.53, 0.75) | 2284.09 |
| *Out-degree, dropping those who name 10 friends* | 0.00 (0.00, 0.39) | 0.44 (0.09, 0.52) | 0.56 (0.46, 0.67) | 1996.58 |

*Note:* 95% confidence intervals indicated in parentheses beneath each estimate. First four models based on 307 monozygotic and 248 dizygotic same-sex twin pairs. The last model drops subjects who named the maximum 10 possible friends, yielding 256 monozygotic and 204 dizygotic same-sex twin pairs. Network measures were transformed to have zero mean and unit variance within each school network to prevent differences between schools from influencing the results.

Since the variance components are not directly observable, the ACE model's assumption of additivity cannot be tested and more complicated relationships are possible. For example, it is possible that genes interact with the environment (GxE) or with other genes (GxG) to yield variation in behavior, or at a higher level, phenotypes interact with the environment (PxE).[13] We limit our analysis to the ACE model but point out that if a strong effect for genes is found in the additive model, then genes are also likely to play a role in more complex specifications as well.

Finally, it is important to clarify the difference between the common environment (C) and the unshared environment (E) in the twin model. Common environment includes the family environment in which both twins were raised, as well as any other factor to which both twins were equally exposed. In contrast, the unshared environment includes idiosyncratic influences that are experienced individually. It is possible to have unshared environmental exposure as a child (e.g., twins may have different friends with different beliefs) and to have shared environments as an adult (e.g., twins may share the same friend). Thus, the distinction between common and unshared environment does not correspond directly to family-nonfamily or adult-child differences in factors that influence a given behavior. Moreover, there may be a similarity in the objective environment but twins may have idiosyncratic experiences that influence their effective environment, and these idiosyncratic experiences may create an unshared rather than a common environmental influence on variation in the phenotype.[13]

**Methods for generating networks from extant models**

*Erdos-Renyi*: we assume there are $N$ nodes and $E$ edges. The probability of a social tie from $i$ to $j$ is $E/(N(N-1))$.[14]

*Fitness*: we assume there are $N$ nodes and $E$ edges. A node $i$ is drawn at random and added at each integer time $t$, each with a fitness $h_i \sim \text{Uniform}[0,1]$. The probability the new node $i$ attaches to any existing node $j$ is $\left(E/N\right)\left(d_j h_j \middle/ \sum_{m=1}^{t} d_m h_m\right)$, where $d_j$ is the degree of node $j$ at the time $i$ is added to the network.[15] Note we do not treat $t$ in this model as an intrinsic characteristic. This means that when we implement the mirror network method, we do not force each twin in a pair to enter the network at the same time $t$.

*Social Space*: we assume there are $N$ nodes and $E$ edges. Each node is placed in a one dimensional social space on the unit interval with uniformly distributed probability. The locations $h_i$ of node $i$ and $h_j$ of node $j$ determine the probability of a mutual social tie between them that equals $1\middle/\left(1+\left(\left|h_i-h_j\right|/\beta\right)^{\alpha}\right)$. The parameters $\alpha = 1.45$ and $\beta = 0.00115$ were chosen to generate the empirically observed mean degree ($E/N$) and transitivity.[16]

*ERGM*: we assume there are $N$ nodes and $E$ edges. We assume a node characteristic $h_i \sim \text{Uniform}[0,1]$ is correlated with in-degree, and we assume the number of transitive triplets is the number observed in the network. Using the ergm function in the STATNET package,[17] we constrain the model to have the observed number of nodes and edges and we choose coefficients for the in-degree characteristic and for the generation of transitive triplets that matches the observed transitivity and heritability of in-degree as closely as possible to the observed transitivity and heritability in the Add Health networks (in our optimization, we find that a coefficient of 2.5 for the in-degree and 2.7 for the transitive triplets performs best). We then simulate networks from this model, keeping the distribution of node characteristics fixed across all simulated networks.[18]

For additional details on how these models were implemented and specific results see the code below.

# Code for "Attract and Introduce" model and "Mirror Network" method

```
# R code for attract and introduce model
# using a mirror network method to measure
# the extent to which this model generates
# heritability of network characteristics
#
# NOTE: Many thanks to Referee 2 who wrote
# some of this code

rm(list=ls())

library(igraph) # Thanks to Gabor Csardi for the igraph package!

n<-750 # number of nodes
e<-3150 # number of edges
m<-e/n # average number of edges per node (degree)

alpha<-0.9
beta<-0.3

T<-10000 # number of simulations

twindeg1 <- rep(0, times=T)
twoutdeg1 <- rep(0, times=T)
twtr1 <- rep(0, times=T)
twbe1 <- rep(0, times=T)

twindeg2 <- rep(0, times=T)
twoutdeg2 <- rep(0, times=T)
twtr2 <- rep(0, times=T)
twbe2 <- rep(0, times=T)

for(t in 1:T) { # do simulation run

  if(t%%100==0) print(t)

  # randomly choose one individual in each network to be a twin
  twin1 <- sample(n,1)
  twin2 <- sample(n,1)

  # create first set of individual traits
  d1<-as.numeric(runif(n)<alpha)*runif(n) # distribution of attractiveness
  c1<-as.numeric(runif(n)<beta) # distribution of prob. of introducing friends

  # create original network
  g1<-graph.empty(n) # create empty graph of size n
  while(ecount(g1)<e) { # loop until enough edges are produced
    dyad<-sample(n,2) # choose a random pair
    if(runif(1)<d1[dyad[2]]) { # check if person 1 nominates person 2 as a friend
      if(runif(1)<c1[dyad[1]]) { # check if person 1 introduces person 2 to friends
        ne<-neighbors(g1,dyad[1]-1,mode="out") # get person 1is neighbors
        if(length(ne)>0) { # check if there is at least one existing neighbor

          for(i in 1:length(ne)) { # loop through neighbors
            if(runif(1)<d1[ne[i]+1]) g1<-add.edges(g1,c(dyad[2]-1,ne[i]))   # introduce friends
              if(runif(1)<d1[dyad[2]]) g1<-add.edges(g1,c(ne[i],dyad[2]-1)) # and add edges if
                                                                           # they become friends
          }
        }
      }
    }
    g1<-add.edges(g1,dyad-1) # add random pair edge at end
  }
  g1<-simplify(g1) # remove duplicate edges and loops
}

  # create second set of individual traits
  d2<-as.numeric(runif(n)<alpha)*runif(n) # distribution of attractiveness
  c2<-as.numeric(runif(n)<beta) # distribution of prob. of introducing friends
```

```
# copy genes
   d2[twin2] <- d1[twin1]
   cl2[twin2] <- cl1[twin1]

# create mirror network
   g2<-graph.empty(n)
   while(ecount(g2)<e) {
      dyad<-sample(n,2)
      if(runif(1)<d2[dyad[2]]) {
         if(runif(1)<cl2[dyad[1]]) {
            ne<-neighbors(g2,dyad[1]-1,mode="out")
            if(length(ne)>0) {
               for(i in 1:length(ne)) {
                  if(runif(1)<d2[ne[i]+1]) g2<-add.edges(g2,c(dyad[2]-1,ne[i]))
                     if(runif(1)<d2[dyad[2]]) g2<-add.edges(g2,c(ne[i],dyad[2]-1))
               }
            }
         }
         g2<-add.edges(g2,dyad-1)
      }
      g2<-simplify(g2)
   }

   # generate/store node statistics for original network
   twindeg1[t] <- degree(g1,v=twin1-1,mode="in")
   twoutdeg1[t] <- degree(g1,v=twin1-1,mode="out")
   twtr1[t] <- transitivity(g1,v=twin1-1,type="local")
   twbe1[t] <- betweenness(g1,v=twin1-1,directed=F)

   # generate/store node statistics for mirror network
   twindeg2[t] <- degree(g2,v=twin2-1,mode="in")
   twoutdeg2[t] <- degree(g2,v=twin2-1,mode="out")
   twtr2[t] <- transitivity(g2,v=twin2-1,type="local")
   twbe2[t] <- betweenness(g2,v=twin2-1,directed=F)
}

corindeg <- cor.test(twindeg1, twindeg2)
coroutdeg <- cor.test(twoutdeg1, twoutdeg2)
cortr <- cor.test(twtr1, twtr2, use="complete.obs")
corbe <- cor.test(twbe1, twbe2)

# Results obtained after 10000 simulations:
corindeg   # 0.46 (0.44,0.48)
coroutdeg  # 0.12 (0.10,0.14)
cortr      # 0.48 (0.46,0.50)
corbe      # 0.29 (0.27,0.31)
```

## Code for "Fitness" model and "Mirror Network" method

```r
# R code for fitness model
# using a mirror network method to measure
# the extent to which this model generates
# heritability of network characteristics
#
# NOTE: Many thanks to Referee 2 who wrote
# some of this code

rm(list=ls())

library(igraph) # Thanks to Gabor Csardi for the igraph package!

n<-750 # number of nodes
e<-3150 # number of edges
m<-e/n # average number of edges per node (degree)

T<-10000 # number of simulations

twindeg1 <- rep(0, times=T)
twoutdeg1 <- rep(0, times=T)
twtr1 <- rep(0, times=T)
twbe1 <- rep(0, times=T)

twindeg2 <- rep(0, times=T)
twoutdeg2 <- rep(0, times=T)
twtr2 <- rep(0, times=T)
twbe2 <- rep(0, times=T)

for(t in 1:T) { # do simulation run

  if(t%%100==0) print(t)

  # randomly choose one individual to be a twin
  twin1 <- sample(n,1)
  twin2 <- sample(n,1)

  # create first set of individual traits
  d1<-runif(n) # distribution of fitness

  # create original network
  g1<-graph.empty(n)
  g1<-add.edges(g1,c(0,1,1,0)) # initialize network
  for(i in 2:(n-1)) {
    p<-degree(g1,v=0:(i-1))*d1[1:i]
    p<-m*p/sum(p)
    ins<-which(runif(i)<p)
    if(length(ins)>0) g1<-add.edges(g1,c(rbind(i,ins-1)))
    g1<-simplify(g1)
  }

  # create second set of individual traits
  d2<-runif(n) # distribution of fitness

  # copy genes
  d2[twin2] <- d1[twin1]

  # create mirror network
  g2<-graph.empty(n)
  g2<-add.edges(g2,c(0,1,1,0))
  for(i in 2:(n-1)) {
    p<-degree(g2,v=0:(i-1))*d2[1:i]
    p<-m*p/sum(p)
    ins<-which(runif(i)<p)
    if(length(ins)>0) g2<-add.edges(g2,c(rbind(i,ins-1)))
    g2<-simplify(g2)
  }
```

```
# generate/store node statistics for original network
  twindeg1[t] <- degree(g1,v=twin1-1,mode="in")
  twoutdeg1[t] <- degree(g1,v=twin1-1,mode="out")
  twtr1[t] <- transitivity(g1,v=twin1-1,type="local")
  twbe1[t] <- betweenness(g1,v=twin1-1,directed=F)

  # generate/store node statistics for mirror network
  twindeg2[t] <- degree(g2,v=twin2-1,mode="in")
  twoutdeg2[t] <- degree(g2,v=twin2-1,mode="out")
  twtr2[t] <- transitivity(g2,v=twin2-1,type="local")
  twbe2[t] <- betweenness(g2,v=twin2-1,directed=F)

}

corindeg <- cor.test(twindeg1, twindeg2)
coroutdeg <- cor.test(twoutdeg1, twoutdeg2)
cortr <- cor.test(twtr1, twtr2, use="complete.obs")
corbe <- cor.test(twbe1, twbe2)

# Results obtained after 10000 simulations:
corindeg    # 0.05 (-0.03,0.01)
coroutdeg   # 0.00 (-0.02,0.02)
cortr       # 0.06 ( 0.04,0.08)
corbe       # 0.02 ( 0.00,0.04)
```

## Code for "Social Space" model and "Mirror Network" method

```r
# R code for social space model
# using a mirror network method to measure
# the extent to which this model generates
# heritability of network characteristics
#
# NOTE: Many thanks to Referee 2 who wrote
# some of this code

rm(list=ls())

library(igraph) # Thanks to Gabor Csardi for the igraph package!

n<-750 # number of nodes
e<-3150 # number of edges
m<-e/n # average number of edges per node (degree)

T<-10000 # number of simulations

alpha<-1.45
beta<-0.00115

twindeg1 <- rep(0, times=T)
twoutdeg1 <- rep(0, times=T)
twtr1 <- rep(0, times=T)
twbe1 <- rep(0, times=T)

twindeg2 <- rep(0, times=T)
twoutdeg2 <- rep(0, times=T)
twtr2 <- rep(0, times=T)
twbe2 <- rep(0, times=T)

for(t in 1:T) { # do simulation run

  if(t%%100==0) print(t)

  # randomly choose one individual in each network to be a twin
  twin1 <- sample(n,1)
  twin2 <- sample(n,1)

  # create first set of individual traits
  d1<-runif(n) # distribution of distances

  # create original network
  dhh1<-abs(outer(d1,d1,"-"))
  rhh1<-1/(1+(dhh1/beta)^alpha)
  A1<-runif(n^2)<rhh1
  diag(A1)<-F
  g1<-simplify(graph.adjacency(A1,mode="upper"))

  # create second set of individual traits
  d2<-runif(n) # distribution of distances

  # copy genes
  d2[twin2] <- d1[twin1]

  # create mirror network
  dhh2<-abs(outer(d2,d2,"-"))
  rhh2<-1/(1+(dhh2/beta)^alpha)
  A2<-runif(n^2)<rhh2
  diag(A2)<-F
  g2<-simplify(graph.adjacency(A2,mode="upper"))

  # generate/store node statistics for original network
  twindeg1[t] <- degree(g1,v=twin1-1,mode="in")
  twoutdeg1[t] <- degree(g1,v=twin1-1,mode="out")
  twtr1[t] <- transitivity(g1,v=twin1-1,type="local")
  twbe1[t] <- betweenness(g1,v=twin1-1,directed=F)
```

```
# generate/store node statistics for mirror network
  twindeg2[t] <- degree(g2,v=twin2-1,mode="in")
  twoutdeg2[t] <- degree(g2,v=twin2-1,mode="out")
  twtr2[t] <- transitivity(g2,v=twin2-1,type="local")
  twbe2[t] <- betweenness(g2,v=twin2-1,directed=F)

}

corindeg <- cor.test(twindeg1, twindeg2)
coroutdeg <- cor.test(twoutdeg1, twoutdeg2)
cortr <- cor.test(twtr1, twtr2, use="complete.obs")
corbe <- cor.test(twbe1, twbe2)

# Results obtained after 10000 simulations:
corindeg     # 0.01 (-0.01,0.03)
coroutdeg    # 0.01 (-0.01,0.03)
cortr        # 0.00 (-0.02,0.02)
corbe        # 0.00 (-0.02,0.02)
```

# Code for Exponential Random Graph Model ("ERGM") and "Mirror Network" method

```r
# R code for ERGM
# using a mirror network method to measure
# the extent to which this model generates
# heritability of network characteristics
#
# NOTE: Many thanks to Referee 2 who wrote
# some of this code

rm(list=ls())

library(ergm)
library(network)

n<-750 # number of nodes
e<-3150 # number of edges
m<-e/n # average number of edges per node (degree)

T<-10000 # number of simulations

twindeg1 <- rep(0, times=T)
twoutdeg1 <- rep(0, times=T)
twtr1 <- rep(0, times=T)
twbe1 <- rep(0, times=T)

twindeg2 <- rep(0, times=T)
twoutdeg2 <- rep(0, times=T)
twtr2 <- rep(0, times=T)
twbe2 <- rep(0, times=T)

for(t in 1:T) { # do simulation run

  if(t%%100==0) print(t)

  # randomly choose one individual in each network to be a twin
  twin1 <- sample(n,1)
  twin2 <- sample(n,1)

  # create first set of individual traits
  d1<-runif(n) # distribution of attractiveness

  # create original network
  g1.use <- network(as.matrix(cbind(sample(n,e,replace=T),sample(n,e,replace=T))))
  g1.use %v% "attract" <- d1 # assign intrinsic characteristic to each node
  g1 <- simulate(~ nodeicov("attract")+ # in degree proportional to intrinsic characteristics
        triadcensus(c(8,11:15)), # count number of transitive triplets
        theta0=c(2.5,rep(2.7,6)), # coefficients 2.5 (for in-degree) and 2.7 (for triplets)
        constraints = ~ edges, # constrain the network to yield e edges
        basis=g1.use, # use basis network to get total number of edges and nodes
        burnin=100000) # let Monte Carlo Chain run a long time before sampling a network

  # create second set of individual traits
  d2<-runif(n) # distribution of attractiveness

  # copy genes to one individual in new network
  d2[twin2] <- d1[twin1]

  # create mirror network
  g2.use <- network(as.matrix(cbind(sample(n,e,replace=T),sample(n,e,replace=T))))
  g2.use %v% "attract" <- d2 # assign intrinsic characteristic to each node
  g2 <- simulate(~ nodeicov("attract")+ # in degree proportional to intrinsic characteristics
        triadcensus(c(8,11:15)), # count number of transitive triplets
        theta0=c(2.5,rep(2.7,6)), # coefficients 2.5 (for in-degree) and 2.7 (for triplets)
        constraints = ~ edges, # constrain the network to yield e edges
        basis=g2.use, # use basis network to get total number of edges and nodes
        burnin=100000) # let Monte Carlo Chain run a long time before sampling a network
```

```
# convert networks to igraph graph objects
  m1 = g1[,]
  m2 = g2[,]

  require(igraph)

  g1i <- graph.adjacency(m1)
  g2i <- graph.adjacency(m2)

  # generate/store node statistics for original network
  twindeg1[t] <- degree(g1i,v=twin1-1,mode="in")
  twoutdeg1[t] <- degree(g1i,v=twin1-1,mode="out")
  twtr1[t] <- transitivity(g1i,v=twin1-1,type="local")
  twbe1[t] <- betweenness(g1i,v=twin1-1,directed=F)

  # generate/store node statistics for mirror network
  twindeg2[t] <- degree(g2i,v=twin2-1,mode="in")
  twoutdeg2[t] <- degree(g2i,v=twin2-1,mode="out")
  twtr2[t] <- transitivity(g2i,v=twin2-1,type="local")
  twbe2[t] <- betweenness(g2i,v=twin2-1,directed=F)

  detach(package:igraph)
}

corindeg <- cor.test(twindeg1, twindeg2)
coroutdeg <- cor.test(twoutdeg1, twoutdeg2)
cortr <- cor.test(twtr1, twtr2, use="complete.obs")
corbe <- cor.test(twbe1, twbe2)

# Results obtained after 10000 simulations:
corindeg    # 0.50  (0.48,0.52)
coroutdeg   # 0.12  (0.10,0.14)
cortr       # 0.02  (0.00,0.04)
corbe       # 0.34  (0.32,0.36)
```

**Fingerprint procedure for assessing the probability of a given distribution of 3-motifs and 4-motifs**

The structure of social networks can be characterized by the distribution of *k-motifs*, or isomorphic combinations of ties between all sets of *k* nodes.[19-22] In a directed network, there are 16 possible combinations of social ties among 3 nodes and 218 possible combinations of social ties among 4 nodes. Several procedures have been proposed for identifying significant motifs,[19-22] but our goal here is to use the motifs to determine which proposed model is most likely to generate a network like the one observed. To do this, we simulate 100 networks from each proposed model using the empirical distribution of nodes and edges in each of the largest 100 Add Health networks (we restrict attention to the largest 100 networks to minimize noise that results from an inadequate number of observations in the smaller networks). We then count the total number of motifs of each type in each network and divide by the total number of motifs in that network to generate the empirical probability that *k* nodes form any given motif (the motif "fingerprint" of the network). For each motif, we calculate the mean ($\mu$) and variance ($\sigma^2$) of the motif probability across all 100 simulated networks. We then use the mean and variance to estimate the $\alpha$ and $\beta$ parameters of one dimension of a multivariate beta density that characterizes the distribution of motif probabilities: $\alpha = \mu^2 (1 - \mu) / \sigma^2 - \mu$;

$\beta = \mu (1 - \mu)^2 / \sigma^2 - (1 - \mu)$. The likelihood of observing a given motif probability can then be estimated from the value of a beta distribution with parameters $\alpha$ and $\beta$ at the point of the observed motif probability. The likelihood of observing a full set of motifs is the product of the likelihoods for each possible motif.

It is important to note that we make a strong assumption with this method that the observed motif probabilities are independent of one another. However, the procedure shows excellent discriminatory power in spite of this strong assumption. We simulated 100 networks from each of the proposed models and used the fingerprint likelihood method we have described to test whether the procedure assigned the highest likelihood to the model that generated the simulated network. Fig. S1 shows the results. The label at the left indicates which model was used to generate the simulated networks, the left panel shows the likelihoods for the 3-motif fingerprint and the right panel shows the likelihoods for the 4-motif fingerprint. For ease of exposition, we show adjusted likelihoods $-\log(c - LL)$, where *c* is a constant across all networks and models and *LL* is the log likelihood of generating an observed network. Each point in the figure represents the adjusted likelihood that a proposed model generated the simulated network.

In all 100 cases for each observed network and for each motif structure, the model with the highest likelihood was the one that generated the data. We also show the likelihood generated by the set of 100 observed networks as well for comparison. We repeated this exercise 10 times (only 1 repetition shown), generating 10 repetitions x 2 fingerprints x 100 networks x 5 models = 10,000 tests of the procedure, and successfully identified the model that generated the network model in all 10,000 cases.

**Figure S1. Fit of Each Model to Simulated Networks**

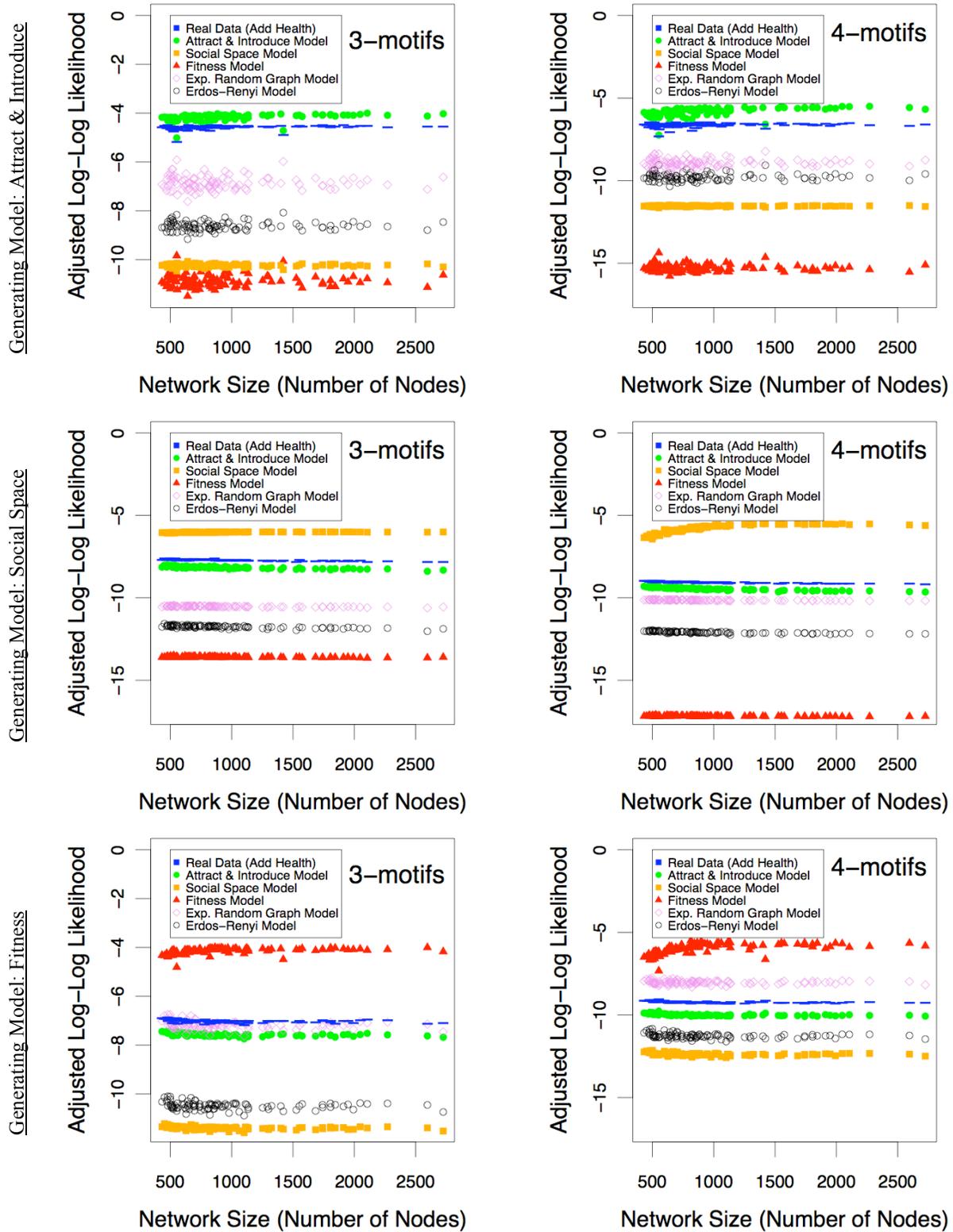

**Figure S1 (continued). Fit of Each Model to Simulated Networks**

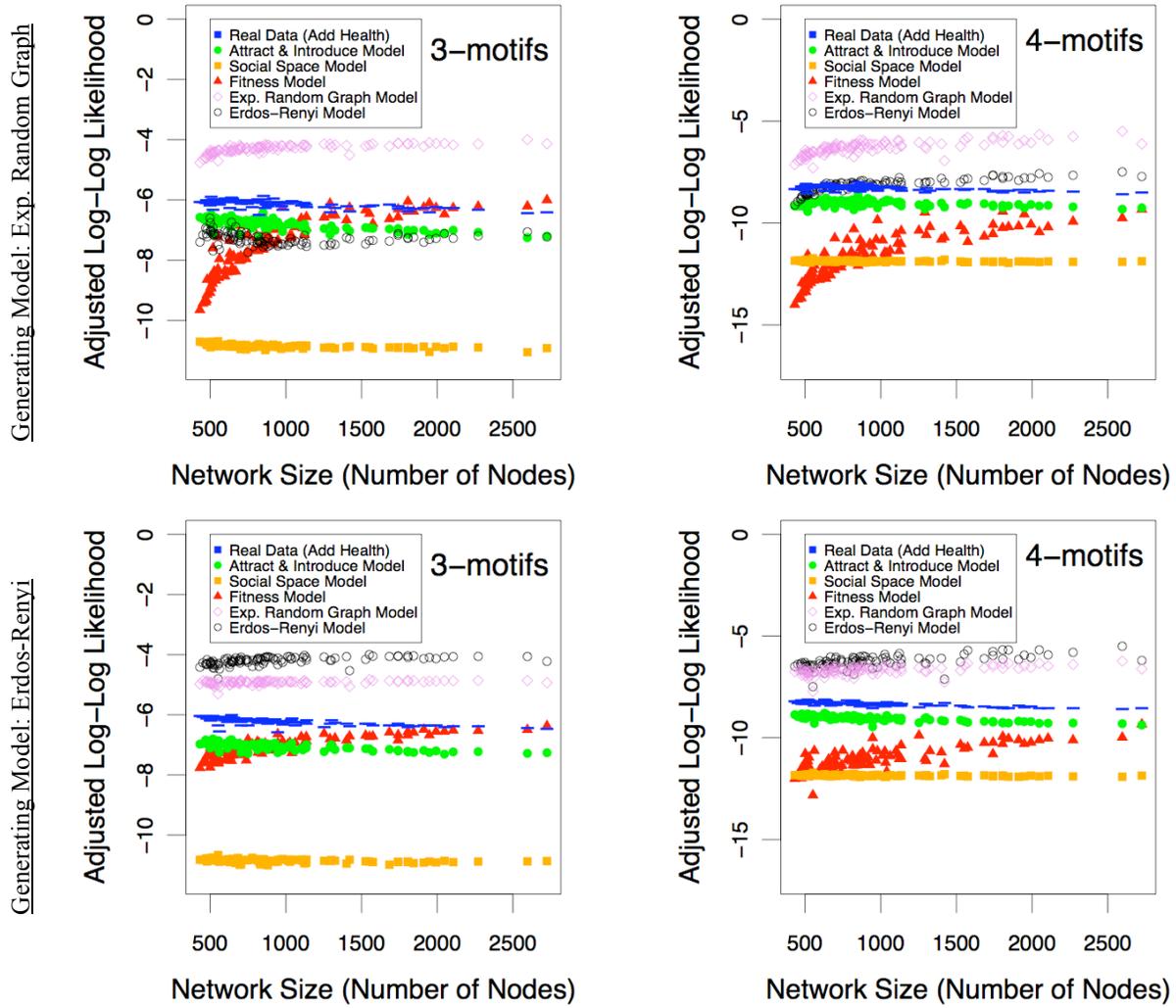

**References for Supporting Information**

1. Udry, J. R.  The National Longitudinal Study of Adolescent Health (Add Health), Waves I & II, 1994–1996; Wave III, 2001–2002 [machine-readable data file and documentation]. Chapel Hill, NC: Carolina Population Center, University of North Carolina at Chapel Hill (2003)

2. Jacobson, KC, and DC Rowe. "Genetic and Shared Environment Influences on Adolescent BMI: Interaction with Race and Sex." *Behavior Genetics* 28:265-275 (1998)

3. Kendler, K.S., Nick G. Martin, A.C. Heath, and L.J. Eaves.  1995.  "Self-report psychiatric symptoms in twins and their nontwin relatives: are twins different? *American Journal of Medical Genetics (Neuropsychiatric Genetics)* 60: 588-591.

4. Bouchard, T. J. 1998. "Genetic and Environmental Influences on Adult Intelligence and Special Mental Abilities." *Human Biology* 70 (2):257-279.

5. Visscher, P.M., Medland, S.E., Ferreira, M.A., Morley, K.I., Zhu, G., Cornes, B.K., Montgomery, G.W., Martin, N.G.  Assumption-free estimation of heritability from genome-wide identity-by-descent sharing between full siblings.  *PLoS Genet* 2:e41 (2006)

6. Bouchard, T. J., and M. McGue. 2003. "Genetic and Environmental Influences on Human Psychological Differences." *Journal of Neurobiology* 54 (1):4-45

7. Posner, S., L.A. Baker, and N.G. Martin. 1996. "Social Contact and Attitude Similarity in Australian Twins." *Behavior Genetics* 26:123-134

8. Harris, Kathleen Mullan, Carolyn Tucker Halpern, Andrew Smolen, and Brett C. Haberstick. 2006. "The National Longitudinal Study of Adolescent Health (Add Health) Twin Data." *Twin Research and Human Genetics* 9 (6):988-997

9. Horwitz, AV, TM Videon, and MF Schmitz. 2003. "Rethinking Twins and Environments: Possible Social Sources for Assumed Genetic Influences in Twin Research." *Journal of Health and Social Behavior* 44 (2):111-129

10. Freese, Jeremy, and Brian Powell. 2003. "Tilting at Twindmills: Rethinking Sociological Responses to Behavioral Genetics." *Journal of Health and Social Behavior* 44 (2):130-135

11. Evans, D.M., Gillespie, N.A., Martin, N.G. *Biological Psychology* 61, 33 (2002).

12. Neale, M.C., Cardon, L.R. *Methodology for Genetic Studies of Twins and Families.* *Dordrecht*, The Netherlands: Kluwer (1992).

13. Bouchard, T.J., Lykken, D.T., McGue, M., Segal, N.L., Tellegen, A.  Sources of human psychological differences: the Minnesota Study of Twins Reared Apart.  *Science* 250: 223 (1990).

14. Erdős, P.; Rényi, A. On Random Graphs. I. *Publicationes Mathematicae* 6: 290-297 (1959)

15. Bianconi, G., Barabási, A.L. Competition and multiscaling in evolving networks.  *Europhys. Lett.* 54: 436 (2001).

16. Boguñá, M., Pastor-Satorras, R., Díaz-Guilera, A., Arenas, A. Models of social networks based on social distance attachment.  *Phys Rev E* 70: 056122 (2004).

17. Handcock, M.S., Hunter, D.R., Butts, C.T., Goodreau, S.M., Morris, M.  ergm: A Package to Fit, Simulate and Diagnose Exponential-Family Models for Networks. Statnet Project, Seattle, WA. Version 2  (2003)

18. Snijders, T.A.B., Pattison, P.E., Robins, G.L., Handcock, M.S. New specifications for exponential random graph models. *Sociological Methodology* 36: 99-153 (2006)

19. Middendorf, M., Ziv, E., Wiggins, C.H.  Inferring network mechanisms. *PNAS* 102: 3192-3197 (2005).

20. Milo, R., Shen-Orr, S., Itzkovitz, S., Kashtan, N., Chklovskii, D., Alon, U.  Network Motifs: Simple Building Blocks of Complex Networks. *Science* 298: 824-827 (2002)

21. Holland, P., Leinhardt, S. in *Sociological Methodology*, D. Heise, Ed. (Jossey-Bass, San Francisco, 1975), pp. 1-45.

22. Wasserman, S., Faust, K.  *Social Network Analysis*. Cambridge Univ. Press, New York, (1994)